\documentclass[reprint,nofootinbib,superscriptaddress,amsmath,amssymb,aps,prd]{revtex4-2}
\usepackage{amsmath,mathrsfs,amsbsy,color,graphicx,bm,amsthm,amsfonts}
\usepackage{bbm}
\usepackage{times}%默认字体为New Times
\usepackage{units}%排版用的
\usepackage{dcolumn}%表格中对齐十进制数
\usepackage{graphicx}%图片插入\right) 
\usepackage{epsfig}
\usepackage{epstopdf}
\usepackage[colorlinks,linkcolor=red,anchorcolor=green,citecolor=blue,CJKbookmarks=True]{hyperref}
\DeclareMathSymbol{\shortminus}{\mathbin}{AMSa}{"39}
\usepackage{mathrsfs}% Align table columns on decimal point
\usepackage{braket}%狄拉克符号
\usepackage{amssymb}%AMS数学符号扩展库
\usepackage{txfonts}%Times字体
\usepackage{enumitem}%提供高度可定制的列表
\usepackage{multirow}%创建复杂表格
\usepackage{orcidlink}
\usepackage{upgreek}
\usepackage{etoolbox}
\newcommand{\recentbibcolor}[1]{}
\AtBeginEnvironment{thebibliography}{\let\savedbibitem\bibitem\renewcommand{\bibitem}[2][]{\recentbibcolor{#2}\savedbibitem[#1]{#2}}}
\begin{document}
\title{Thermodynamics of rotating black holes with constrained primary hair}

\author{Wentao Liu\orcidlink{0009-0008-9257-8155}}
\email[]{liuwentao@lzu.edu.cn} 
\affiliation{Lanzhou Center for Theoretical Physics, Key Laboratory of Theoretical Physics of Gansu Province, 
Key Laboratory of Quantum Theory and Applications of MoE,
Gansu Provincial Research Center for Basic Disciplines of Quantum Physics, 
Lanzhou University, Lanzhou 730000, China}
\affiliation{Institute of Theoretical Physics $\&$ Research Center of Gravitation,
Lanzhou University, Lanzhou 730000, China}

\author{Di Wu\orcidlink{0000-0002-2509-6729}}
\email[]{wdcwnu@163.com}
\affiliation{School of Physics and Astronomy, China West Normal University, Nanchong, Sichuan 637002, P. R. China}
\affiliation{Department of Physics and Astronomy, University of Waterloo, Waterloo, Ontario N2L 3G1, Canada}

\author{Shao-Wen Wei\orcidlink{0000-0003-0731-0610}}
\email[]{weishw@lzu.edu.cn} 
\affiliation{Lanzhou Center for Theoretical Physics, Key Laboratory of Theoretical Physics of Gansu Province, 
Key Laboratory of Quantum Theory and Applications of MoE, Gansu Provincial Research Center for Basic Disciplines of Quantum Physics, 
Lanzhou University, Lanzhou 730000, China}
\affiliation{Institute of Theoretical Physics $\&$ Research Center of Gravitation,
Lanzhou University, Lanzhou 730000, China}

\author{Yu-Xiao Liu\orcidlink{0000-0002-4117-4176}}
\email[]{liuyx@lzu.edu.cn (Corresponding author)} 
\affiliation{Lanzhou Center for Theoretical Physics, Key Laboratory of Theoretical Physics of Gansu Province, 
Key Laboratory of Quantum Theory and Applications of MoE, Gansu Provincial Research Center for Basic Disciplines of Quantum Physics, 
Lanzhou University, Lanzhou 730000, China}
\affiliation{Institute of Theoretical Physics $\&$ Research Center of Gravitation,
Lanzhou University, Lanzhou 730000, China}

\begin{abstract}
An exact rotating geometry need not define a black hole in equilibrium when its surface gravity varies across the horizon. 
We construct an asymptotically flat, non-Kerr subfamily of the aligned null solutions of generalized Proca theory whose outer horizon is exactly Killing and isothermal to second order in rotation. 
The horizon conditions constrain the vector hair, and integrability of the normalized horizon charge fixes the remaining second-order amplitude. 
Computing the complete gravitational and nonminimal vector surface charges, we obtain the asymptotically normalized mass and total angular momentum, and derive a logarithmic Hamiltonian entropy. 
The first law holds in both independent directions of the resulting two-parameter solution space. 
We express the mass as a function of entropy and angular momentum, recover the geometric thermodynamic potentials, and establish a scale-corrected Smarr relation that retains the dimensionful coupling. 
The exact metric defines the geometry, while equilibrium thermodynamics is consistently restricted to second order in spin.
\end{abstract}

\maketitle
\raggedbottom

\section{Introduction}

An exact rotating black-hole solution does not by itself determine a thermodynamic state space \cite{Ma:2026reducedphase,Wei:2026topologyreview}. 
The field equations determine the geometry and matter configuration, whereas a uniform horizon temperature and an integrable entropy charge place further requirements on the solution family and its allowed variations \cite{Liu:2014scalarcharges}. 
These requirements become particularly consequential when nonminimally coupled hair introduces integration data beyond mass and angular momentum \cite{Liu:2025oho}. 
Recent beyond-Horndeski solutions exhibit an independent primary scalar-hair parameter \cite{Bakopoulos:2024primaryscalar}, and their Euclidean thermodynamics has been developed explicitly \cite{Bakopoulos:2024hairthermo}. 
These results motivate examining how independent hair enters the thermodynamic variations in other nonminimally coupled theories.
Thermodynamic constructions for rotating Lovelock black branes, scalar-haired black holes, and higher-derivative AdS black holes illustrate how additional interactions enter this connection between geometry and thermodynamics \cite{Dehghani:2006branes,Erices:2018hairy,Hu:2024reallsantos}, while canonical integrability can also determine the energy and conjugate potentials when the asymptotic frame is nontrivial \cite{Zhang:2026knbr}. 
Here we ask how horizon conditions and charge integrability can be satisfied simultaneously within an exact rotating family with vector hair.

Generalized Proca theory provides a concrete setting for this question \cite{Heisenberg:2014rta,Geng:2016einsteinvector,Dong:2024procapolarizations,Lai:2024einsteinvector,Dong:2025vectortensor}. 
Its regularized Gauss--Bonnet realization admits static black holes with primary vector hair \cite{Charmousis:2025proca}; related constructions yield regular black holes and gravastars \cite{Eichhorn:2025regularproca}, a tower of Lovelock--Proca interactions with static hairy solutions \cite{Fernandes:2025lovelockproca}, and effective cosmological constants arising as integration data \cite{Charmousis:2026effectivehair}. 
Fernandes's exact rotating solutions extend this setting through an aligned null Kerr--Schild ansatz \cite{Fernandes:2026primary}. 
Their vector integration data deform the geometry and generically break circularity. 
The geometric consequences of relaxing circularity have also been analyzed systematically, with explicit examples separating null horizons from Killing horizons \cite{Babichev:2025circularity,Anson:2021disformal}. 
This freedom makes explicit the distinction between solving the field equations and constructing an equilibrium horizon: a null surface need not be a Killing horizon, and the usual matter-regularity assumptions that ensure constant surface gravity need not hold. 
The exact family thus allows the horizon requirements to be imposed directly on its integration data.

We retain this exact geometric description and use a spin expansion to control the horizon and thermodynamic conditions. 
Slow-rotation methods have been developed for higher-curvature theories and nontrivial matter couplings, supporting both solution construction and thermodynamic analyses \cite{Pani:2011slow,Adair:2020slow,Marks:2023slow,Gray:2022lense,Kubiznak:2022nle,Yue:2011slow,Stetsko:2019slow}. 
Expanding thermodynamic quantities from an exact geometry is also useful, as in the grand canonical treatment of Kerr--Newman \cite{Chen:2001slowKN} and the slow-rotation equation of state following from Kerr--AdS thermodynamics \cite{Gibbons:2005firstlaw,Gunasekaran:2012extended,Altamirano:2014rotating,Wei:2016kerrcritical}, including its application to thermodynamic chaos \cite{Tang:2021chaos}. 
In the present problem, the expansion serves a further purpose: it resolves two successive restrictions on the hair. 
A constant-radius Killing horizon and uniform surface gravity through second order fix the angular data and leading amplitude. 
Requiring the normalized horizon charge to be integrable then selects the remaining second-order amplitude at fixed theory couplings.

The resulting two-parameter subfamily connects an exact non-Kerr geometry to an integrable Hamiltonian entropy. 
We first establish the thermodynamics of its exact static member and then derive the second-order rotational corrections. 
Using the solution-space charge prescription and the complete generalized Proca boundary terms \cite{Hajian:2016solutionphase,Minamitsuji:2024thermodynamics}, we obtain the asymptotically normalized mass, angular momentum, and a logarithmic entropy that differs from one quarter of the horizon area. 
The fundamental relation reproduces the geometric temperature and angular velocity, and its scaling identity retains the dimensionful coupling. 
The restriction on the hair has a concrete charge-theoretic signature: allowing the second-order amplitude to vary independently makes the same normalized horizon charge nonintegrable. 
Our construction identifies an integrable subfamily for this charge prescription, with the static deformation kept finite and rotational equilibrium controlled through second order.
The remainder of this paper is organized as follows.
Section \ref{sec:background} constructs the rotating branch and establishes its horizon properties.
Section \ref{sec:static} presents its exact static member and derives the static
thermodynamics. Section \ref{sec:charges} extends this construction to slow
rotation, including the canonical charges, horizon integrability, fundamental
relation, and scaling identity. 
Finally, Sec. \ref{sec:conclusions} summarizes our main findings and discusses their broader implications and future applications.
%We conclude in Sec. \ref{sec:conclusions}.

\section{Solutions and horizon conditions}\label{sec:background}

\subsection{Generalized Proca theory and exact rotating solutions}

Generalized Proca theory extends the Einstein-Proca system by derivative self-interactions of a massive vector field $A_\mu$, arranged so that the vector sector propagates three physical polarizations \cite{Heisenberg:2014rta}. 
We consider the following $G_2$-$G_4$ sector:
\begin{equation}\label{Action}
\mathcal S=\frac{1}{16\pi}\int d^4x\sqrt{-g}\,
(\mathcal L_2+\mathcal L_3+\mathcal L_4),
\end{equation}
where
\begin{equation*}
\begin{aligned}
\mathcal L_2&=G_2(X),\\
\mathcal L_3&=G_3(X)\nabla_\mu A^\mu,\\
\mathcal L_4&=G_4(X)\mathcal R+G_{4,X}(X)\left[
(\nabla_\mu A^\mu)^2-\nabla_\rho A_\sigma\nabla^\sigma A^\rho
\right].
\end{aligned}
\end{equation*}
Here $X\equiv-\frac{1}{2}A_\mu A^\mu$, $\mathcal R$ is the Ricci scalar, $G_{i,X}\equiv dG_i/dX$, and we use $G=c=\hbar=k_{\rm B}=1$.
The action \eqref{Action} also contains the four-dimensional Proca theories obtained by regularizing Lovelock invariants \cite{Charmousis:2025proca,Fernandes:2025lovelockproca}. 
Up to a common normalization, the $n$-th Lovelock density generates
\begin{equation}\label{LovelockProcaFunctions}
\begin{aligned}
G_2^{(n)}(X)&=-6\,2^n(n-1)X^n,\\
G_3^{(n)}(X)&=-2^{n+1}nX^{n-1},\\
G_4^{(n)}(X)&=-\frac{2^{n-1}n}{2n-3}X^{n-1}.
\end{aligned}
\end{equation}
For $n=2$, Eq. \eqref{LovelockProcaFunctions} reproduces the Proca theory associated with the four-dimensional regularized Gauss-Bonnet invariant; higher Lovelock densities supply an infinite tower of related interactions. 
The interactions in Eq. \eqref{LovelockProcaFunctions}, supplemented by the Einstein--Hilbert term and their coupling constants, belong to a subset of generalized Proca theory \cite{Charmousis:2025proca}.
Here we consider the broader class of functions with the following expansions near $X=0$:
\begin{equation}\label{GPfunctions}
\begin{aligned}
G_2(X)&=O(X^2),\\
G_3(X)&=\alpha c_{3,1}X+O(X^2),\\
G_4(X)&=1+\alpha c_{4,1}X+\alpha^2c_{4,2}X^2
+O(X^3).
\end{aligned}
\end{equation}
Here $\alpha$ has dimensions of length squared, whereas $c_{3,1}$, $c_{4,1}$, and $c_{4,2}$ are dimensionless couplings. 
For the aligned null configurations considered below, $X=0$, and only the coefficients displayed in Eq. \eqref{GPfunctions} enter the field equations.

The null ansatz implements the same structural idea that makes the Kerr-Newman solution tractable \cite{Newman:1965my}: the metric is written in Kerr-Schild form and the vector field is aligned with the Kerr-Schild null congruence. 
In ingoing Kerr coordinates $(v,r,\theta,\tilde{\varphi})$, we use the metric ansatz
\begingroup
\medmuskip=2mu plus 1mu minus 1mu
\begin{equation}\label{primaryKerrSchild}
\begin{aligned}
ds^2={}&\!-\!\left[1\!-\!\frac{2rM(r,\theta)}{\Sigma }\right]
\left(dv+a\sin^2\theta\,d\tilde{\varphi}\right)^2
+\Sigma\sin^2\theta\,d\tilde{\varphi}^{\,2}\\
&+\Sigma\, d\theta^2
+2\left(dv+a\sin^2\theta\,d\tilde{\varphi}\right)
\left(dr+a\sin^2\theta\,d\tilde{\varphi}\right).
\end{aligned}
\end{equation}
\endgroup
Here, $\Sigma=r^2+a^2\cos^2\theta$ and $a$ is the rotation parameter. 
The Proca field is aligned with the Kerr-Schild null congruence according to
\begin{equation}
A_\mu dx^\mu=
\frac{\mathcal{A}(r,\theta)}{\Sigma}
\left(dv+a\sin^2\theta\,d\tilde{\varphi}\right).
\end{equation}
The alignment makes the vector field null, $X=0$.
For $c_{4,2}=-c_{4,1}^2/2$ with $c_{3,1}\neq0$ and $c_{4,1}\neq0$, the exact rotating solution given in Sec.~4.1 of Ref.~\cite{Fernandes:2026primary} can be written using the dimensionless auxiliary function
\begin{equation}\label{auxiliaryChi}
\chi(r,\theta)=3\arctan\frac{a\cos\theta}{r}
-\frac{ar\cos\theta(3r^2+5a^2\cos^2\theta)}{\Sigma^2}.
\end{equation}
The functions $\mathcal A(r,\theta)$ and $M(r,\theta)$ are then
\begin{align}
\mathcal A(r,\theta)
&=q(\theta)\left[1+
\frac{\alpha c_{4,1}^2 q(\theta)\chi(r,\theta)}{2c_{3,1}a^3\cos^3\theta}\right]^{-1},
\label{FernandesExactVector}\\[4pt]
M(r,\theta)
&=m+\frac{\alpha c_{4,1}q(\theta)^2}{2r\Sigma}
\left[1+\frac{\alpha c_{4,1}^2 q(\theta)\chi(r,\theta)}{2c_{3,1}a^3\cos^3\theta}\right]^{-2}.
\label{FernandesExactMass}
\end{align}
Here $q(\theta)$ is an arbitrary angular integration function, and $m$ is the mass integration constant, denoted by $\mu$ in Ref.~\cite{Fernandes:2026primary}.
For the Kerr-Schild ansatz \eqref{primaryKerrSchild}, circularity requires $M(r,\theta)=M(r)$; any nontrivial angular dependence of the generalized mass function thus describes a noncircular rotating spacetime.

\subsection{A rotating branch with an equilibrium outer horizon}

We first reparametrize the exact solution in Eqs.~\eqref{FernandesExactVector} and \eqref{FernandesExactMass} and express the metric as the standard Kerr geometry plus a correction due to the vector hair. 
We use the same coupling conditions and, for definiteness, take the functions in Eq.~\eqref{GPfunctions} to terminate at the displayed orders, with $G_2=0$. 
Define
\begin{equation}\label{hairparameter}
k=\frac{4\alpha c_{4,1}^2}{c_{3,1}},\qquad
\gamma=\frac{\alpha c_{4,1}}{k^2},
\end{equation}
where $k$ and $\gamma$ have dimensions of length squared and inverse length squared, respectively. In the subfamily selected below, $\gamma$ controls the hair-induced departure from Kerr, while $k$ sets the overall amplitude of the vector field. They depend only on the theory couplings and are held fixed in the thermodynamic variations.

To rewrite the exact solution in a compact form, we set $z=a\cos\theta$ and define, for $r>0$,
\begin{equation}\label{radialPrimitive}
\mathcal F(r,z)=\int_r^\infty\frac{ds}{(s^2+z^2)^3},\qquad
\mathcal F(r,0)=\frac1{5r^5}.
\end{equation}
For $z\neq0$, this integral evaluates to
\begin{equation}\label{radialPrimitiveExplicit}
\mathcal F(r,z)=\frac{3}{8z^5}\arctan\frac zr
-\frac{3r}{8z^4\Sigma}-\frac{r}{4z^2\Sigma^2}.
\end{equation}
The value at $z=0$ in Eq.~\eqref{radialPrimitive} is its continuous limit, covering both the equatorial plane and the static case.

In terms of these quantities, $\mathcal A$ and $M$ take the equivalent form
\begin{align}
\mathcal A(r,\theta)&=\bigl[\mathcal C(\theta)+kz^2\mathcal F(r,z)\bigr]^{-1},
\label{generalVectorProfile}\\
M(r,\theta)&=m+\frac{\gamma k^2\mathcal A^2}{2r\Sigma}.
\label{generalMassFunction}
\end{align}
Here $\mathcal C(\theta)=1/q(\theta)$ wherever $q(\theta)\neq0$. The angular function remains arbitrary and will be constrained by the horizon conditions below. The constant $m$ is the ADM mass when $\mathcal A$ approaches a finite function at infinity; the mass function $M(r,\theta)$ is distinct from the total mass $M$ used in the thermodynamic analysis.

In Kerr-adapted coordinates $(t,r,\theta,\varphi)$, with $\Delta_{\rm Kerr}=r^2-2mr+a^2$, the aligned null one-form is
\begin{equation}\label{adaptedNullForm}
\ell=dt-a\sin^2\theta\,d\varphi
+\frac{\Sigma}{\Delta_{\rm Kerr}}\,dr.
\end{equation}
The full metric and vector field are then
\begin{equation}\label{ds2}
ds^2=ds^2_{\rm Kerr}+\frac{\gamma k^2\mathcal A^2}{\Sigma^2}\ell^2,
\qquad A_\mu dx^\mu=\frac{\mathcal A}{\Sigma}\ell.
\end{equation}
Here $ds^2_{\rm Kerr}$ is the standard Kerr metric in Boyer--Lindquist coordinates \cite{Boyer:1966qh}. For the Killing-horizon family constructed below, the function $\mathcal A(r,\theta)$ is given explicitly in Eq.~\eqref{horizonAdaptedProfile}, subject to the horizon relation \eqref{horizonMass}. The term $\gamma k^2\mathcal A^2\ell^2/\Sigma^2$ gives the vector-hair correction with the full dependence on $a$ retained. For members for which $\mathcal A(r,\theta)$ approaches a finite angular function at infinity, the deformation coefficient in Eq.~\eqref{ds2} decays as $r^{-4}$ and the geometry is asymptotically flat. The formal zero-deformation limit restores the Kerr metric \cite{Kerr:1963ud}. The angular dependence of $M$ preserves the generic noncircularity of the full geometry in these Kerr-adapted coordinates. In particular, $\Delta_{\rm Kerr}=0$ is a coordinate singularity introduced by the coordinate transformation; the horizon must be determined from the full metric. The equivalence of the expressions for $\mathcal A$ and $M$ and the explicit metric components are detailed in Appendix~\ref{app:rotatinggeometry}, where Eq.~\eqref{KerrCoordinateTransformation} gives the coordinate transformation.

The preceding reparametrization leaves $\mathcal C(\theta)$ arbitrary. We now restrict this exact solution family to members admitting a constant-radius Killing horizon. In regular ingoing coordinates, a stationary axisymmetric null surface $r=H(\theta)$ obeys
\begin{equation}\label{deformedHorizon}
[H'(\theta)]^2+\Delta(H(\theta),\theta)=0,\qquad
\Delta=r^2+a^2-2rM.
\end{equation}
For a constant-radius surface $r=R$, Eq.~\eqref{deformedHorizon} reduces to $\Delta(R,\theta)=0$. By Eq.~\eqref{generalMassFunction}, this requires $\mathcal A^2(R,\theta)/\Sigma_H$ to be independent of angle. We therefore set
\begin{equation}\label{horizonVectorData}
\mathcal A(R,\theta)=b\sqrt{\Sigma_H},\qquad
\Sigma_H=R^2+a^2\cos^2\theta,
\end{equation}
where $b$ is an angle-independent constant. For given $R$, $a$, and $b$, this condition determines $\mathcal C(\theta)$ through Eq.~\eqref{generalVectorProfile}. Substituting Eqs.~\eqref{generalMassFunction} and \eqref{horizonVectorData} into Eq.~\eqref{deformedHorizon} with $H(\theta)=R$ gives
\begin{equation}\label{horizonMass}
R^2-2mR+a^2-\gamma k^2b^2=0.
\end{equation}
Using the horizon value \eqref{horizonVectorData} to eliminate $\mathcal C(\theta)$ from Eq.~\eqref{generalVectorProfile}, we obtain
\begin{equation}\label{horizonAdaptedProfile}
\mathcal A=
\left\{\frac1{b\sqrt{\Sigma_H}}
-kz^2[\mathcal F(R,z)-\mathcal F(r,z)]\right\}^{-1},
\end{equation}
where $b\neq0$ remains a free parameter, with $m$, $a$, $R$, and $b$ satisfying the horizon relation \eqref{horizonMass}. Substitution into Eq.~\eqref{ds2} gives the corresponding exact solution.

These data give $\Delta(R,\theta)=0$ exactly. 
To identify the null surface as a Killing horizon, consider
\begin{equation}\label{horizonGenerator}
\chi_H=\partial_t+\Omega_H\partial_\varphi,\qquad
\Omega_H=\frac{a}{R^2+a^2}.
\end{equation}
The vector $\chi_H$ is null and normal to $r=R$, establishing the Killing-horizon property without assuming circularity. The direct verification is given in Appendix~\ref{app:horizonexterior}.

The horizon conditions leave the amplitude $b$ free. We next constrain it by requiring uniform surface gravity through second order in rotation \cite{Bardeen:1973fourlaws,Biswas:2025procazeroth}. The exact surface gravity of this Killing horizon is
\begin{equation}\label{exactSurfaceGravity}
\begin{aligned}
\kappa_H(\theta)&=\frac{\partial_r\Delta|_R}{2(R^2+a^2)}
=\frac{R-m-\gamma k^2b^2 L_H(\theta)}{R^2+a^2},\\
L_H(\theta)&=\frac{kbz^2}{\Sigma_H^{5/2}}-\frac R{\Sigma_H}.
\end{aligned}
\end{equation}
For a branch even in $a$, write $b=b_0+O(a^2)$. 
Its angular expansion is
\begin{equation}
L_H=-\frac1R+\frac{a^2\cos^2\theta}{R^3}
\left(1+\frac{kb_0}{R^2}\right)+O(a^4).
\end{equation}
For nonzero hair and $\gamma\neq0$, uniform surface gravity through second order therefore selects
\begin{equation}\label{thermalHairRoot}
b_0=-\frac{R^2}{k}.
\end{equation}
The leading hair amplitude is therefore tied to the horizon radius.

Uniform surface gravity leaves the second-order monopole amplitude undetermined. 
We write this remaining freedom as
\begin{equation}\label{secondOrderAmplitude}
b=-\frac{R^2}{k}\left[1+u(R)\frac{a^2}{R^2}\right]+O(a^4),
\qquad y=\gamma R^2.
\end{equation}
Thus Eqs.~\eqref{ds2}, \eqref{horizonMass}, and \eqref{horizonAdaptedProfile} describe the exact family with a Killing horizon. Equation~\eqref{secondOrderAmplitude} enforces uniform surface gravity through $a^2$, while $u(R)$ remains undetermined. We retain this freedom in the surface-charge calculation and determine it from entropy integrability in Sec.~\ref{sec:entropy}.

The entropy requires a separate charge calculation. 
The nonzero horizon value $A_\mu\chi_H^\mu=b\sqrt{\Sigma_H}/(R^2+a^2)$ prevents a direct use of the usual smooth-bifurcation-surface reduction \cite{Wald:1993nt,Iyer:1994noether}. 
The entropy must therefore be obtained from the complete Hamiltonian surface charge, rather than identified with one quarter of the horizon area by assumption. The static member is presented in Sec.~\ref{sec:static}.

\section{Static black hole and thermodynamics}\label{sec:static}
\subsection{The exact static solution}

The static member isolates the effect of vector hair before spin corrections enter. 
Setting $a=0$ in Eqs.~\eqref{horizonMass} and \eqref{horizonAdaptedProfile}, with $b_0$ fixed by Eq.~\eqref{thermalHairRoot}, gives $\mathcal A=q_0=-R^3/k$. 
In regular ingoing coordinates the complete solution takes the simple form
\begin{equation}\label{staticIngoingSolution}
\begin{aligned}
ds^2&=-f(r)dv^2+2\,dv\,dr+r^2d\Omega_2^2,\\
A&=\frac{q_0}{r^2}dv,\qquad
f(r)=1-\frac{2M_0}{r}-\frac{\gamma R^6}{r^4},\\
M_0&=\frac R2(1-\gamma R^2),\qquad q_0=-\frac{R^3}{k}.
\end{aligned}
\end{equation}
The exterior time coordinate $\tau$, defined by $dv=d\tau+dr/f(r)$, diagonalizes the metric:
\begin{equation}\label{staticDiagonalSolution}
\begin{aligned}
ds^2&=-f(r)d\tau^2+\frac{dr^2}{f(r)}+r^2d\Omega_2^2,\\
A&=\frac{q_0}{r^2}\left(d\tau+\frac{dr}{f(r)}\right).
\end{aligned}
\end{equation}
The radial component of $A$ arises from the change of coordinates; the field remains null, $X=0$, and is regular at the future horizon in the ingoing chart \eqref{staticIngoingSolution}.
The $r^{-4}$ correction cannot be absorbed into the mass and gives $\mathcal R=6\gamma R^6/r^6$, distinguishing this geometry from the Schwarzschild vacuum solution.
The constraint $q_0=-R^3/k$, inherited from the rotating equilibrium branch, leaves $R$ as the sole independent parameter at fixed $\gamma$ and $k$; it is not imposed by static spherical symmetry alone.
For $0<-\gamma R^2\leq0.1$, $r=R$ is a nonextremal outer Killing horizon generated by $\partial_v=\partial_\tau$, with surface gravity and temperature
\begin{equation}\label{staticTemperature}
\kappa_0=\frac{f'(R)}2=\frac{1+3y}{2R},\qquad
T_0=\frac{1+3y}{4\pi R},\qquad y=\gamma R^2.
\end{equation}
Spherical symmetry ensures that the surface gravity is exactly uniform over the horizon.

\subsection{Static thermodynamics}

With the temperature fixed geometrically, we obtain the energy and entropy from the surface charges. 
Asymptotic normalization selects $\partial_v$ as the time generator.
Using the full generalized Proca surface charge, including its nonminimal vector contribution \cite{Minamitsuji:2024thermodynamics},
Appendix \ref{app:surfacecharge} gives at $a=0$
\begin{equation}\label{staticSurfaceCharge}
\oint\delta\boldsymbol Q_{\partial_v}=\frac{\delta M_0}{2},
\qquad
-\oint\partial_v\cdot\boldsymbol\Theta=\frac{\delta M_0}{2}.
\end{equation}
Thus $\delta H_{\partial_v}=\delta M_0$, independently of the extraction radius.

The Hamiltonian entropy charge of $\partial_v/T_0$, holding the generator fixed in the surface variation, obeys
\begin{equation}\label{staticEntropyVariation}
dS_0=\frac{dM_0}{T_0}
=\frac{2\pi R(1-3y)}{1+3y}\,dR.
\end{equation}
Integrating at fixed couplings and setting the additive constant by the formal $R\to0$ limit gives
\begin{equation}\label{staticThermodynamics}
S_0=-\pi R^2+\frac{2\pi}{3\gamma}\ln(1+3\gamma R^2),
\qquad dM_0=T_0dS_0.
\end{equation}
Both $T_0$ and $S_0$ are positive in the stated domain.
The logarithmic term is part of the Hamiltonian entropy of this constrained family. 
The nonzero $A\cdot\partial_v=q_0/R^2$ at the horizon prevents the usual smooth bifurcation-surface reduction, so an area entropy cannot be assumed.
The formal $\gamma\to0$ limit restores $M_0=R/2$, $T_0=1/(4\pi R)$, and $S_0=\pi R^2$.

Since $S_0'(R)>0$, Eq. \eqref{staticThermodynamics} defines $R$
implicitly as a function of entropy. The static fundamental relation is
\begin{equation}
M_0(S_0;\gamma)=\frac{R(S_0;\gamma)}2
\left[1-\gamma R(S_0;\gamma)^2\right],
\end{equation}
whose entropy derivative gives $(\partial M_0/\partial S_0)_\gamma=T_0$.
Scaling the radius and the dimensionful coupling as
$(R,\gamma)\to(\lambda R,\lambda^{-2}\gamma)$ then yields
\begin{equation}\label{staticScaling}
\begin{aligned}
M_0&=2T_0S_0-2\gamma\Pi_0,\\
\Pi_0&=\left(\frac{\partial M_0}{\partial\gamma}\right)_{S_0}
=\frac{(1+3y)\ln(1+3y)-3y-3y^2}{6\gamma^2R}.
\end{aligned}
\end{equation}
Here $\Pi_0$ expresses a scaling identity across theory couplings;
the physical first law above keeps $\gamma$ and $k$ fixed. It does not
introduce an independently variable hair charge or assert an extended
Hamiltonian first law with coupling variations.

\section{Slowly rotating black hole thermodynamics}\label{sec:charges}
\subsection{Canonical surface charges}

We now compute the mass and angular momentum of the rotating family in Eqs.~\eqref{ds2}, \eqref{horizonMass}, and \eqref{horizonAdaptedProfile} through second order in $a$. At fixed $\gamma$ and $k$, we use the expansion \eqref{secondOrderAmplitude} and take $(R,a)$ as independent parameters for each prescribed function $u(R)$. The function $u(R)$ remains unspecified until the entropy-integrability analysis in Sec.~\ref{sec:entropy}. 
Its tangent variations obey $\delta X=0$ and $\delta b=(\partial_Rb)\delta R+(\partial_ab)\delta a$.
This relation defines the restricted solution space used here; it is not derived from an independent boundary condition fixing the hair.
We use the canonical-integrability strategy of Ref. \cite{Zhang:2026knbr}, with the charges evaluated for the present asymptotically flat Proca theory.
As emphasized by Mann and collaborators \cite{Adair:2020slow,Marks:2023slow}, rotational corrections to entropy and temperature first appear at quadratic order, so their thermodynamic effects require a slow-rotation treatment through at least $O(a^2)$. 
For the present branch, these quantities are likewise even in $a$. 
We therefore retain scalar thermodynamic quantities through $a^2$ and angular momentum and angular velocity through $a$, which captures the leading rotational corrections. 
The resulting thermodynamic one-form has errors $O(a^4)dR+O(a^3)da$, with no expansion in $y=\gamma R^2$.
Accordingly, all rotating thermodynamic relations below are understood perturbatively to the retained order.

Writing the dynamical fields collectively as $\Phi=(g_{\mu\nu},A_\mu)$, the variation of the Lagrangian four-form is \cite{Lee:1990symmetries,Iyer:1994noether}
\begin{equation}
\delta\boldsymbol L=\boldsymbol E\cdot\delta\Phi
+d\boldsymbol\Theta(\Phi,\delta\Phi).
\end{equation}
For an exact symmetry $\xi$, the surface-charge two-form is \cite{Iyer:1994noether}
\begin{equation}\label{surfaceChargeDefinition}
\boldsymbol k_\xi=\delta\boldsymbol Q_\xi
-\xi\cdot\boldsymbol\Theta,
\qquad \delta\xi=0,
\end{equation}
where $\boldsymbol Q_\xi$ is the Noether charge. 
The two-form $\boldsymbol k_\xi$ is closed on shell for a linearized solution. Its integral is therefore independent of the homologous closed two-surface chosen in the regular exterior.

Path independence in solution space instead requires
\begin{equation}\label{chargeIntegrability}
(\delta_1\delta_2-\delta_2\delta_1)H_\xi=0,
\qquad \delta H_\xi=\oint\boldsymbol k_\xi,
\end{equation}
which is the charge-integrability condition \cite{Hajian:2016solutionphase}.
This distinguishes radial conservation of a surface charge from the existence of its state function on the selected family.
The relevant exact symmetries are time translation and axial rotation, which preserve both fields, $\mathcal L_{\partial_t}\Phi=\mathcal L_{\partial_\varphi}\Phi=0$.
The $G_3$ and $G_4$ interactions do not admit an independent transformation $A\rightarrow A+d\lambda$. 
Thus there is no Maxwell gauge charge. This does not exclude a hair work term on a larger solution space; such a term would require its own
boundary-variation analysis \cite{Liu:2014procathermo,Feng:2016horndeskimaxwell}.
In particular, the nonzero horizon value of $A\cdot\chi_H$ is not an electromagnetic chemical potential.

On the null solution space, $G_4=1$ and $G_{4,X}=\gamma k^2$, whereas $G_3=0$ and $\delta X=0$. 
The terms proportional to $\gamma k^2$ in the surface charge remain nonzero and must be included \cite{Minamitsuji:2024thermodynamics}. Evaluating the full charge in regular ingoing coordinates gives
\begin{equation}\label{massAndAngularCharges}
\delta H_{\partial_t}=\delta m,\qquad
\delta H_{\partial_\varphi}=-\delta(ma)
\end{equation}
at the retained orders. The axial sign is chosen so that positive $a$ corresponds to positive angular momentum. With the zero-charge reference fixed by the vanishing-size limit,
\begin{equation}\label{canonicalCharges}
M=m,\qquad J=ma+O(a^3).
\end{equation}
These are the total charges, including the nonminimal vector contribution, not charges assigned to the metric sector alone.

The cancellation between the Noether and boundary-potential terms explains why the full charge is essential. 
Setting $q_0=-R^3/k$, we evaluate the Noether integrals on a sphere of fixed coordinate radius $r$ and obtain
\begin{equation}\label{finiteRadiusNoether}
\begin{aligned}
\oint\boldsymbol Q_{\partial_t}
&=\frac m2-\frac{\gamma k^3q_0^3a^2}{6r^8}+O(a^4),\\
\oint\boldsymbol Q_{\partial_\varphi}&=-ma+O(a^3).
\end{aligned}
\end{equation}
The boundary-potential contribution cancels the $r^{-8}$ term in $\delta H_{\partial_t}$; the details are given in Appendix \ref{app:surfacecharge}. 
Thus the Noether integral alone is not the energy. The conserved quantity is the complete surface variation in Eq. \eqref{surfaceChargeDefinition}.

These integrable charges also fix the canonical frame. 
In the notation of Ref. \cite{Zhang:2026knbr}, a candidate mass generator is
\begin{equation}
\xi_M=\mathcal N\bigl(\partial_t+\Omega_{\rm int}\partial_\varphi\bigr),
\qquad
\delta H_{\xi_M}=\mathcal N(\delta m-\Omega_{\rm int}\delta J).
\end{equation}
The coefficients are held fixed inside the surface variation. 
Here the asymptotic conditions $g_{tt}\rightarrow-1$ and a nonrotating frame at infinity select
\begin{equation}\label{canonicalFrame}
\mathcal N=1,\qquad\Omega_{\rm int}=0.
\end{equation}
The coordinate-time charge is already integrable, so no solution-dependent time normalization is needed. 
The canonical frame is the Kerr-adapted asymptotic frame of Sec. \ref{sec:background}. 
A compensating Maxwell gauge shift, used in the electrovacuum construction, has no counterpart in this Proca model.

\subsection{Horizon integrability and entropy}\label{sec:entropy}

Having fixed the asymptotic charges, we now determine $u(R)$ from entropy integrability and integrate the resulting horizon charge to obtain the entropy. 
Here $T_H=\kappa_H/(2\pi)$ is evaluated from Eqs.~\eqref{exactSurfaceGravity} and \eqref{secondOrderAmplitude} through $a^2$, retaining the undetermined $u(R)$ \cite{Hawking:1975vcx}.
The normalized horizon generator depends on the solution through $\Omega_H$ and $T_H$. 
We implement the fixed-generator surface variation explicitly:
\begin{equation}\label{adjustedChargePrescription}
\begin{aligned}
\delta^{[\Phi]}\boldsymbol Q_\eta
&\equiv\delta\boldsymbol Q_\eta-\boldsymbol Q_{\delta\eta},\\
\boldsymbol k_\eta[\delta\Phi;\Phi]
&=\delta^{[\Phi]}\boldsymbol Q_\eta-\eta\cdot\boldsymbol\Theta.
\end{aligned}
\end{equation}
Here $\delta$ is the total parameter variation, whereas $\delta^{[\Phi]}$ varies the dynamical fields at fixed symmetry label.
For $\eta=c_t\partial_t+c_\varphi\partial_\varphi$, with coefficients constant in spacetime but dependent on $(R,a)$, linearity gives
\begin{equation}
\boldsymbol Q_{\delta\eta}
=(\delta c_t)\boldsymbol Q_{\partial_t}
+(\delta c_\varphi)\boldsymbol Q_{\partial_\varphi}.
\end{equation}
These terms cancel the variations of the coefficients in $\delta\boldsymbol Q_\eta$, so that
\begin{equation}\label{fieldDependentLinearity}
\oint\boldsymbol k_\eta=c_t\,dM-c_\varphi\,dJ.
\end{equation}

Applying this prescription to $\chi_H=\partial_t+\Omega_H\partial_\varphi$ gives
\begin{equation}\label{horizonChargeFirstLaw}
\oint\boldsymbol k_{\chi_H}=dM-\Omega_HdJ.
\end{equation}
For the normalized generator $\eta_H=\chi_H/T_H$, the Hamiltonian entropy prescription \cite{Hajian:2016solutionphase} gives the candidate entropy variation
\begin{equation}\label{entropyChargeDefinition}
\omega\equiv\oint\boldsymbol k_{\eta_H}
=\frac{dM-\Omega_HdJ}{T_H}.
\end{equation}
Thus no $dT_H$ or $d\Omega_H$ terms enter the surface variation itself.
The coefficients still vary when testing closure of this one-form on solution space.

The resulting one-form defines an entropy only if it is integrable. With $u(R)$ still arbitrary in Eq.~\eqref{secondOrderAmplitude}, the charges above and the geometric temperature give
\begin{equation}\label{entropyIntegrability}
\partial_a\omega_R-\partial_R\omega_a
=\frac{16\pi\gamma Ra}{(1+3y)^2}
\big[(3y-1)u-1\big]+O(a^3).
\end{equation}
Thus integrability fixes $u=(3y-1)^{-1}$ at the retained order. The one-form is then locally integrable at this order, and we define $dS_H=\omega$. This additional condition selects a thermodynamic subfamily of the geometries constructed in Sec.~\ref{sec:background}; the field equations and horizon conditions alone leave $u$ undetermined. The independent second-order hair direction has a nonzero charge curl, as shown in Appendix~\ref{app:independenthair}.

Substituting $u=(3y-1)^{-1}$ into the charges and horizon quantities gives the expansions entering Eq.~\eqref{entropyChargeDefinition}:
\begin{equation}\label{thermodynamicExpansion}
\begin{aligned}
&M=M_0+a^2M_2+O(a^4),\\
&M_0=\frac R2(1-y),\qquad
M_2=\frac{1-y}{2R(1-3y)},\\
&T_H=T_0+a^2T_2+O(a^4),\\
&T_0=\frac{1+3y}{4\pi R},\qquad
T_2=-\frac{(1+3y)(2-3y)}{4\pi R^3(1-3y)},\\
&J\,=aM_0+O(a^3),\qquad
\Omega_H=\frac a{R^2}+O(a^3).
\end{aligned}
\end{equation}

The entropy must extend the static result. 
Writing $S_H=S_0+a^2S_2+O(a^4)$, the zeroth-order radial variation and the leading spin variation give, respectively,
\begin{equation}\label{entropyCoefficients}
S_0'=\frac{M_0'}{T_0}
=\frac{2\pi R(1-3y)}{1+3y},\qquad
S_2=\frac{2M_2-M_0/R^2}{2T_0}.
\end{equation}
Primes denote derivatives with respect to $R$ at fixed $\gamma$.
The first equation recovers the static entropy variation; the second fixes its spin correction. 
Together they give
\begin{equation}\label{HamiltonianEntropy}
\begin{aligned}
S_H={}&-\pi R^2+\frac{2\pi}{3\gamma}\ln(1+3\gamma R^2)\\
&+\pi a^2\frac{1-\gamma R^2}{1-3\gamma R^2}+O(a^4).
\end{aligned}
\end{equation}
We choose the additive constant so that $S_H$ tends to zero in the formal $R\rightarrow0$ limit with bounded $a/R$. 

The remaining radial coefficient checks consistency of the integrated entropy,
\begin{equation}\label{radialFirstLawCheck}
M_2'-\frac{M_0'}{R^2}=T_0S_2'+T_2S_0',
\end{equation}
which is satisfied identically. 
At fixed couplings, the perturbative first-law relation takes the form
\begin{equation}\label{firstLaw}
dM=T_HdS_H+\Omega_HdJ +O(a^3).
\end{equation}
Here $O(a^3)$ denotes the remainder of the thermodynamic one-form, understood componentwise in $(R,a)$.
The radial check verifies internal consistency of the integrated horizon charge in both parameter directions; the first law follows from this charge construction.

Hamiltonian treatments of Horndeski and generalized Proca black holes likewise retain the contributions of nonminimal matter couplings when extracting entropy variations \cite{Feng:2015horndeskientropy,Feng:2016horndeskimaxwell,Minamitsuji:2023horndeski,Minamitsuji:2024thermodynamics}.
The entropy in Eq. \eqref{HamiltonianEntropy} differs from $A_H/4=\pi(R^2+a^2)$. 
This difference originates in the complete horizon charge of the nonminimally coupled field \cite{Liu:2026nonwald}. 
The horizon contraction $A\cdot\chi_H\neq0$ precludes the usual smooth extension of the stationary vector to a bifurcation surface, so the reduction to an area term is not available. 
We define $S_H$ by the solution-space Hamiltonian prescription \cite{Hajian:2016solutionphase}, using the boundary representative in Appendix \ref{app:surfacecharge}. 
At fixed extraction radius, the parametric perturbations solve the linearized equations and the exact symmetry makes $\boldsymbol k_{\eta_H}$ closed. 
Its finite integral can therefore be evaluated on exterior spheres and continued to a regular future-horizon section, without a bifurcation surface. 
The complete $-\eta_H\cdot\boldsymbol\Theta$ term is retained in this limit.
This establishes the Lorentzian charge definition on the selected family. 
A conventional smooth Euclidean saddle \cite{Gibbons:1977euclidean} does not supply an independent entropy derivation here: already at $a=0$, the continued vector has $\oint A_E=-2\pi i(q_0/R^2)/\kappa_0\ne0$ around a shrinking thermal circle, incompatible with a smooth one-form at its center.

The thermodynamic results above can be realized by an exact member of the original solution family. We choose the following completion of $b$ with the required second-order expansion:
\begin{equation}\label{exactAmplitudeChoice}
C_b=1+\frac{a^2}{R^2(1-3y)},\qquad
b=-\frac{R^2}{kC_b}.
\end{equation}
This completion is not unique beyond second order. Any smooth even completion with the same expansion through $a^2$ gives the same thermodynamics at the retained order.

Combining Eqs. \eqref{horizonAdaptedProfile} and
\eqref{exactAmplitudeChoice}, define
\begin{equation}\label{finalRadialFunction}
U_b=\frac{C_b}{R^2\sqrt{\Sigma_H}}
+z^2[\mathcal F(R,z)-\mathcal F(r,z)].
\end{equation}
With this choice of exact completion, the second-order equilibrium
branch has the exact representative
\begin{equation}\label{selectedExactBranch}
\begin{aligned}
ds^2=ds^2_{\rm Kerr}+\frac{\gamma}{\Sigma^2U_b^2}\ell^2,\quad\quad A_\mu dx^\mu=-\frac1{k\Sigma U_b}\ell.
\end{aligned}
\end{equation}
Here the Kerr mass parameter $m$ is fixed by the horizon condition \eqref{horizonMass}, after substituting $b=-R^2/(kC_b)$.
At fixed $\gamma$ and $k$, its independent parameters are $(R,a)$.

For this representative, a sufficient domain in which $R$ bounds a regular, nonextremal outer horizon is
\begin{equation}\label{admissibleParameterRegion}
k<0,\quad \gamma<0,\quad
0<-\gamma R^2\leq0.1,\quad 0<\frac{|a|}{R}\leq0.1.
\end{equation}
The exterior estimates are given in Appendix~\ref{app:horizonexterior}.
The logarithm in Eq.~\eqref{HamiltonianEntropy} is real in the domain \eqref{admissibleParameterRegion}, where the entropy and temperature are positive.
The metric \eqref{selectedExactBranch} remains exact, whereas equilibrium thermodynamics is restricted to second order in $a/R$: angular variations of $\kappa_H$ start at fourth order.

\subsection{Fundamental relation and scaling}\label{sec:thermodynamics}

The first law can now be checked in thermodynamic coordinates by expressing the mass in terms of $(S_H,J)$. 
To invert the logarithmic entropy, introduce a radius $\rho$ through
\begin{equation}\label{entropyRadius}
S_H=S_0(\rho;\gamma),\qquad
S_0(\rho;\gamma)=-\pi\rho^2+
\frac{2\pi}{3\gamma}\ln(1+3\gamma\rho^2).
\end{equation}
Since $S_0'>0$ in the chosen domain, this relation is locally invertible.
At nonzero spin $\rho$ is not the geometric horizon radius:
\begin{equation}
\rho=R+a^2\frac{S_2(R)}{S_0'(R)}+O(a^4).
\end{equation}

Using this inversion to eliminate $R$ and $a$ from Eq. \eqref{thermodynamicExpansion} gives the fundamental relation
\begin{equation}\label{fundamentalRelation}
M(S_H,J;\gamma)=\frac{\rho}{2}(1-\gamma\rho^2)
+\frac{J^2}{\rho^3(1-\gamma\rho^2)}+O(J^4),
\end{equation}
where $\rho$ is fixed implicitly by Eq. \eqref{entropyRadius}. 
This is a slow-rotation expansion at finite $\gamma\rho^2$, not an expansion in the hair strength. 
In deriving it, the useful cancellation is $M_2-T_0S_2=M_0/(2R^2)$. 
The nonzero-$\gamma$ relation is determined by these charges and the logarithmic entropy; imposing the Kerr Christodoulou--Ruffini formula \cite{Christodoulou:1971mass} would instead assume the area law.

To recover the conjugate potentials, set $c(\rho)=[\rho^3(1-\gamma\rho^2)]^{-1}$ and differentiate the fundamental relation at fixed coupling:
\begin{equation}\label{canonicalPotentials}
\begin{aligned}
T&=\left(\frac{\partial M}{\partial S_H}\right)_{J,\gamma}
=\frac{M_0'(\rho)+J^2c'(\rho)}{S_0'(\rho)}+O(J^4),\\
\Omega&=\left(\frac{\partial M}{\partial J}\right)_{S_H,\gamma}
=\frac{2J}{\rho^3(1-\gamma\rho^2)}+O(J^3).
\end{aligned}
\end{equation}
Transforming back to $(R,a)$ reproduces
\begin{equation}\label{potentialMatching}
T=T_H+O(a^4),\qquad \Omega=\Omega_H+O(a^3).
\end{equation}
Thus the thermodynamic derivatives agree with the geometric horizon potentials, and Eq. \eqref{canonicalFrame} is consistent with the first law. 
There is no independent Proca chemical potential on this constrained solution space.

The same fundamental relation determines the effect of the coupling scale. 
Dimensionful parameters contribute to scaling relations in extended black-hole thermodynamics
\cite{Gunasekaran:2012extended,Poshteh:2021scaling,Cheng:2016kncritical,Wei:2014gbtriple,Wei:2020gbextended,Wei:2022gblandscape,Chen:2023massiveholography,Chen:2025variableNewton}.
Here $\gamma$ enters the scaling identity while remaining fixed in the physical first law. 
Consider the simultaneous scaling
\begin{equation}
(R,a,\gamma)\longrightarrow(\lambda R,\lambda a,\lambda^{-2}\gamma).
\end{equation}
With the entropy normalization in Eq. \eqref{HamiltonianEntropy}, the charges scale as $(S_H,J,M)\rightarrow(\lambda^2S_H,\lambda^2J,\lambda M)$.
To keep track of the coupling scale, define the response of the computed state function across this family of theories,
\begin{equation}\label{couplingResponse}
\Pi_\gamma\equiv
\left(\frac{\partial M}{\partial\gamma}\right)_{S_H,J}.
\end{equation}
Euler's identity then yields the scale-corrected Smarr relation
\begin{equation}\label{SmarrRelation}
M=2T_HS_H+2\Omega_HJ-2\gamma\Pi_\gamma+O(a^4).
\end{equation}
The additional term $-2\gamma\Pi_\gamma$ arises from the dependence of the mass on the dimensionful coupling $\gamma$, whose scaling weight is $-2$. 
It therefore contributes to the Euler scaling relation, while the corresponding variation is absent from the fixed-coupling first law \eqref{firstLaw}.
Evaluating the coupling derivative gives
\begin{equation}\label{explicitCouplingResponse}
\begin{aligned}
\Pi_\gamma&=\Pi_0
\left[1-\frac{2-3y}{1-3y}\frac{a^2}{R^2}\right]+O(a^4),\\
\Pi_0&=\frac{(1+3y)\ln(1+3y)-3y-3y^2}
{6\gamma^2R}.
\end{aligned}
\end{equation}
Direct substitution verifies Eq. \eqref{SmarrRelation} separately at orders $a^0$ and $a^2$. Omitting its last term would give an incorrect Kerr-type Smarr formula.
The response in Eq. \eqref{couplingResponse} is used here to express a scaling identity. It is not an additional independent hair charge, and
no $d\gamma$ term is included in the fixed-theory first law \eqref{firstLaw}. 
Interpreting it as a work potential in an extended Hamiltonian problem would additionally require varying the action couplings and including their bulk contributions \cite{Kastor:2010lovelockfirstlaw,Xiao:2026thermodynamicvolume}. 
This distinction is essential: the external-field parameter in the reference electrovacuum family labels solutions of one theory, whereas $\gamma=\alpha c_{4,1}/k^2$ labels the theory itself.

Two limits check the construction. Setting $a=0$ recovers the exact static relations of Sec. \ref{sec:static}, including Eq. \eqref{staticThermodynamics}. 
The formal limit $\gamma\rightarrow0$ tests the normalization against Kerr. 
It gives $S_H=\pi(R^2+a^2)+O(a^4)$ and the slow-rotation Kerr mass, temperature, and angular velocity. 
With $\rho=\sqrt{S_H/\pi}$, Eq. \eqref{fundamentalRelation} becomes
\begin{equation}
M=\frac\rho2+\frac{J^2}{\rho^3}+O(J^4),
\qquad M^2=\frac{S_H}{4\pi}+\frac{\pi J^2}{S_H}+O(J^4).
\end{equation}
Also $\gamma\Pi_\gamma\rightarrow0$, restoring the Kerr Smarr relation \cite{Smarr:1972kt}.
This limit checks the formulas across theory parameters; it is not a variation performed in Eq. \eqref{firstLaw}. 
Neither the slow-rotation truncation nor the sufficient domain \eqref{admissibleParameterRegion} supports extrapolation to an extremal endpoint.

\section{Conclusion and discussion}\label{sec:conclusions}

We have constructed a thermodynamic subfamily of the exact rotating primary-hair solutions in generalized Proca theory. 
Its metric retains a standard Kerr part plus a single squared-one-form correction. 
The horizon conditions fix the angular vector data and the leading amplitude; integrability of the normalized horizon charge fixes the second-order monopole correction. 
The resulting family is asymptotically flat and non-Kerr, with an exact Killing outer horizon and uniform surface gravity through second order in rotation.

The canonical surface-charge calculation includes the nonminimal Proca terms. 
The asymptotically normalized time translation already generates an integrable mass, $M=m$, and the total angular momentum is $J=ma$ at the required order. 
The nontrivial integrability condition instead lies in the entropy sector. 
Its solution gives the logarithmic Hamiltonian entropy \eqref{HamiltonianEntropy}, for which the first-law relation holds in both independent parameter directions through second order in spin.
The fundamental relation reproduces the geometric temperature and angular velocity, while its scaling identity contains an explicit contribution from the dimensionful coupling.
The formal Kerr limit recovers both the area entropy and the Christodoulou--Ruffini relation at the retained order, providing a normalization check independent of the finite-coupling scaling identity.

The resulting Hamiltonian first-law structure applies to the specified two-parameter branch through second order in spin. 
The vector hair is constrained, the finite-spin surface gravity acquires angular dependence at fourth order, and the entropy is defined by the complete surface charge rather than a smooth-bifurcation-surface area prescription. 

Beyond second order, one could investigate whether the area-averaged surface gravity supports an effective first law of the form
\[
\delta M=T_{\mathrm{eff}}\,\delta S_{\mathrm{eff}}+\Omega_H\,\delta J,
\]
with the candidate temperature
\[
T_{\mathrm{eff}}\equiv\frac{\kappa_{\mathrm{eff}}}{2\pi},
\qquad
\kappa_{\mathrm{eff}}\equiv\frac{1}{A_H}
\int_{\mathcal H}\kappa_H(\theta)\,dA,
\]
where $\mathcal H$ is a spatial horizon cross-section. This proposal requires testing the integrability of $(dM-\Omega_HdJ)/T_{\mathrm{eff}}$ using the complete Proca surface charge; the physical interpretation of the averaged temperature would also need to be established.

Extending the construction to independent hair variations, establishing a physical-process second law, and studying dynamical stability are distinct questions. 
Gravitational effective field theory also admits entropy constructions satisfying the second law through quadratic order in perturbations \cite{Hollands:2022eft}. Recent dynamical-entropy constructions address first-order departures from stationary horizons \cite{Hollands:2024entropy,Visser:2024entropy}, while a linearized second law has been established for nonminimally coupled vector fields on bifurcate Killing backgrounds \cite{Wall:2024vectorlaw}. Relating these constructions to the present future-horizon charge requires accounting for the vector field's lack of a smooth bifurcation-surface extension. The exact
geometry and the integrable charge derived here provide a definite
starting point for those analyses.

\acknowledgments
The authors thank Dr. Zelin Zhang for helpful discussions. 
This work is supported in part by the National Natural Science Foundation of China under Grant Nos. 12605108, 12547147, 12475056, 12475055, 12675077 and 12247101; the Fundamental Research Funds for the Central Universities under Grant Nos. lzujbky-2024-jdzx06 and lzujbky-2025-jdzx07; Gansu Province’s Top Leading Talent Support Plan; the Natural Science Foundation of Gansu Province under Grant Nos. 22JR5RA389 and 25JRRA799; the ``111 Center'' under Grant No. B20063; the China Postdoctoral Science Foundation under Grant No. 2025M783393; the China Scholarship Council; and the Sichuan Science and Technology Program under Grant No. 2026NSFSC0021.
\appendix

\section{Equivalent forms of the rotating solution}\label{app:rotatinggeometry}

This appendix verifies that Eqs.~\eqref{generalVectorProfile} and \eqref{generalMassFunction} are equivalent to the expressions of Sec.~II\,A, and derives the Kerr-adapted form in Eqs.~\eqref{adaptedNullForm} and \eqref{ds2}. The angular integration function remains arbitrary throughout; no horizon or equilibrium conditions are imposed here.

The function $\mathcal A(r,\theta)$ satisfies the radial equation
\begin{equation}\label{radialProfileEquation}
\partial_r\mathcal A=\frac{k a^2\cos^2\theta}{\Sigma^3}\mathcal A^2.
\end{equation}
Since $\partial_r\mathcal F=-\Sigma^{-3}$, radial integration gives Eq.~\eqref{generalVectorProfile} with an angular integration function $\mathcal C(\theta)$. For $q(\theta)\neq0$, the identity $\chi=8z^5\mathcal F$ rewrites the original expression \eqref{FernandesExactVector} as
\begin{equation}\label{radialProfileEquivalence}
\mathcal A^{-1}=\frac1{q(\theta)}
+\frac{\alpha c_{4,1}^2\chi(r,\theta)}{2c_{3,1}z^3}
=\mathcal C(\theta)+kz^2\mathcal F(r,z).
\end{equation}
Thus $\mathcal C(\theta)=1/q(\theta)$, establishing the equivalence with Eq.~\eqref{generalVectorProfile}. Substituting this expression for $\mathcal A$ and $\gamma k^2=\alpha c_{4,1}$ into Eq.~\eqref{FernandesExactMass} gives Eq.~\eqref{generalMassFunction}. The zero-hair solution follows directly from Eq.~\eqref{FernandesExactVector} by setting $q=0$.

For the Kerr-adapted form \eqref{ds2}, define
\begin{equation}
\begin{aligned}
\Delta_{\rm Kerr}&=r^2-2mr+a^2,\\
\delta M(r,\theta)&=M(r,\theta)-m,
\end{aligned}
\end{equation}
where $\delta M$ denotes the deformation of the mass function. The coordinate transformation is
\begin{equation}\label{KerrCoordinateTransformation}
\begin{aligned}
dv&=dt+\frac{r^2+a^2}{\Delta_{\rm Kerr}}\,dr,\\
d\tilde{\varphi}&=-d\varphi-\frac{a}{\Delta_{\rm Kerr}}\,dr.
\end{aligned}
\end{equation}
The azimuthal sign gives the standard Boyer--Lindquist convention for the Kerr background,
\begin{equation}\label{KerrBL}
\begin{aligned}
ds^2_{\rm Kerr}={}&-\frac{\Delta_{\rm Kerr}}{\Sigma}
\left(dt-a\sin^2\theta\,d\varphi\right)^2
+\frac{\Sigma}{\Delta_{\rm Kerr}}\,dr^2\\
&+\Sigma\,d\theta^2
+\frac{\sin^2\theta}{\Sigma}
\left[a\,dt-(r^2+a^2)d\varphi\right]^2.
\end{aligned}
\end{equation}
The same transformation maps $dv+a\sin^2\theta\,d\tilde\varphi$ to the one-form $\ell$ in Eq.~\eqref{adaptedNullForm}.
Substituting $M=m+\delta M$ and $2r\delta M/\Sigma=\gamma k^2\mathcal A^2/\Sigma^2$ into the original metric gives the decomposition \eqref{ds2}. The full metric \eqref{ds2} has nonvanishing components
\begin{equation}\label{primaryMetricComponents}
\begin{aligned}
g_{tt}&=-1+\frac{2rM}{\Sigma},\\[2pt]
g_{t\varphi}&=-\frac{2arM\sin^2\theta}{\Sigma},\\[2pt]
g_{\varphi\varphi}&=\biggl(r^2+a^2+
\frac{2a^2rM\sin^2\theta}{\Sigma}\biggr)\sin^2\theta,\\[2pt]
g_{rr}&=\frac{\Sigma}{\Delta_{\rm Kerr}}
+\frac{2r\Sigma\delta M}{\Delta_{\rm Kerr}^2},\\[2pt]
g_{\theta\theta}&=\Sigma,\\[2pt]
g_{tr}&=\frac{2r\delta M}{\Delta_{\rm Kerr}},\\[2pt]
g_{r\varphi}&=-\frac{2ar\delta M\sin^2\theta}{\Delta_{\rm Kerr}}.
\end{aligned}
\end{equation}
Here $M$ denotes the mass function $M(r,\theta)$.

\section{Horizon normality and exterior regularity}\label{app:horizonexterior}

We first verify the Killing-horizon property of the family in Eqs.~\eqref{horizonMass} and \eqref{horizonAdaptedProfile}, as used in Sec.~\ref{sec:background}\,B, with $b$ left free.

To verify normality directly, use the regular ingoing chart of the original metric, with the standard azimuthal orientation $\phi=-\tilde\varphi$. 
The Killing vectors satisfy $\partial_v=\partial_t$ and $\partial_\phi=\partial_\varphi$. 
Then $\chi_H=\partial_v+\Omega_H\partial_\phi$, and substitution of $2RM(R,\theta)=R^2+a^2$ into the metric gives
\begin{equation}\label{horizonNormality}
\left.\chi_H^2\right|_R=0,\qquad
\left.\chi_{H\mu}\right|_R
=\frac{\Sigma_H}{R^2+a^2}\,\partial_\mu r.
\end{equation}
Since $\Sigma_H>0$, the Killing vector is a nonvanishing normal to $r=R$, establishing the Killing-horizon property without assuming circularity.

We now turn to the exact representative \eqref{selectedExactBranch} selected in Sec.~\ref{sec:entropy}, with $b$ fixed by Eq.~\eqref{exactAmplitudeChoice}, and verify exterior regularity in the sufficient parameter domain \eqref{admissibleParameterRegion}.
The bound $0.1$ is a convenient choice for the analytic estimates and has no special physical significance; it is not an error tolerance for the slow-rotation expansion.
Here $C_b\geq1$ and $U_b>0$ for $r\geq R$. 
In particular, $0<\mathcal A\leq R^3\sqrt{1.01}/|k|$. Setting $s=-\gamma R^2$, the radial horizon function satisfies the uniform bound
\begin{equation}\label{exteriorMonotonicity}
\begin{aligned}
\frac{\partial_r\Delta}{R}
&\geq0.99-s\left[1+2(1.01)+0.02(1.01)^{3/2}\right]\\
&>0.68,\qquad r\geq R.
\end{aligned}
\end{equation}
Together with $U_b>0$, this excludes radial poles in $\mathcal A$ and further zeros of $\Delta$, and establishes nonextremality.
The causal interpretation follows in regular ingoing coordinates with the standard azimuthal orientation: the future outgoing null direction has
$dr/dv=\Delta/[2(r^2+a^2)]>0$ outside $R$, while $\nabla r$ becomes parallel to the future horizon generator at $R$. 
The constant-radius Killing surface therefore bounds the asymptotically flat exterior.
Equation \eqref{admissibleParameterRegion} is a sufficient domain, not a claim about the maximal parameter range.

\section{Surface charge on the null Proca branch}\label{app:surfacecharge}

We first give the exact boundary potential and Noether charge for the null Proca branch, then evaluate their surface integrals in the slow-rotation expansion.
The variations are taken at fixed $\gamma,k$ and preserve $X=0$, so that $\delta X=0$.
We use the ingoing chart with $\phi=-\tilde\varphi$ and orientation $\epsilon_{vr\theta\phi}=+\sqrt{-g}$, and define $h_{\mu\nu}=\delta g_{\mu\nu}$, $h^{\mu\nu}=g^{\mu\alpha}g^{\nu\beta}h_{\alpha\beta}$, $h=g^{\mu\nu}h_{\mu\nu}$, and $F_{\mu\nu}=2\nabla_{[\mu}A_{\nu]}$.
For the chosen theory, $G_4(0)=1$ and $G_{4,X}(0)=\gamma k^2$; the boundary potential obtained by direct variation \cite{Minamitsuji:2024thermodynamics} is
\begin{equation}
\Theta_{abc}=\frac1{16\pi}\epsilon_{dabc}\,\vartheta^d,
\end{equation}
where
\begingroup
\thinmuskip=2mu
\medmuskip=2mu plus 1mu minus 1mu
\thickmuskip=3mu plus 1mu minus 1mu
\begin{equation}\label{fullBoundaryPotential}
\begin{aligned}
\vartheta^\rho\!=&\nabla_\nu h^{\rho\nu}-\nabla^\rho h
+\gamma k^2 A^\alpha(g^{\nu\rho}\nabla^\sigma A_\alpha
-g^{\sigma\nu}\nabla^\rho A_\alpha)h_{\sigma\nu}\\
&+\gamma k^2(\nabla\cdot A)\left(2g^{\rho\sigma}\delta A_\sigma
-2g^{\rho\sigma}A^\nu h_{\nu\sigma}+A^\rho h\right)\\
&\!-\!2\gamma k^2\big[(\nabla^\sigma A^\rho)\delta A_\sigma\!-A^\nu\nabla^{(\sigma}A^{\rho)}h_{\nu\sigma}+\frac12A^\rho\nabla^\nu A^\sigma h_{\sigma\nu}\big].
\end{aligned}
\end{equation}
\endgroup
The corresponding Noether two-form is
\begin{equation}
(Q_\xi)_{ab}=-\frac1{16\pi}\epsilon_{ab\mu\nu}P_\xi^{\mu\nu},
\end{equation}
with
\begin{equation}\label{fullNoetherKernel}
\begin{aligned}
P_\xi^{\mu\nu}={}&\nabla^\mu\xi^\nu
+2\gamma k^2 A^\alpha\nabla^\mu A_\alpha\,\xi^\nu
-2\gamma k^2(\nabla\cdot A)A^\mu\xi^\nu\\
&-\frac{\gamma k^2}2 F^{\mu\nu}(A\cdot\xi)
+\gamma k^2 A^\mu(\nabla^\nu A_\alpha+\nabla_\alpha A^\nu)\xi^\alpha.
\end{aligned}
\end{equation}
The contraction with $\epsilon_{ab\mu\nu}$ selects the antisymmetric part of $P_\xi^{\mu\nu}$.
These expressions are exact on the null branch; no expansion in the rotation parameter has been made.

The $G_3$ interaction contributes no surface-charge variation on this branch.
In the variation of $G_3(X)\nabla_\mu A^\mu$, the term $G_{3,X}\delta X\,\nabla_\mu A^\mu$ contains no derivatives of the field variations and therefore produces no boundary potential.
The boundary potential and Noether charge from the remaining terms are proportional to $G_3$, while
\begin{equation}
G_3(0)=0,\qquad \delta G_3=G_{3,X}(0)\delta X=0.
\end{equation}
Thus $\boldsymbol\Theta^{(3)}$ and $\delta\boldsymbol Q^{(3)}_\xi$ vanish even though $G_{3,X}(0)\neq0$.
The nonminimal contributions proportional to $\gamma k^2$ in Eqs. \eqref{fullBoundaryPotential} and \eqref{fullNoetherKernel} must still be retained.

We now introduce the slow-rotation expansion to evaluate the charges.
Keeping the second-order amplitude $u(R)$ arbitrary, the vector field takes the form
\begin{equation}\label{boundaryVectorExpansion}
\begin{aligned}
&\!A_v=\frac{q_0}{r^2}+a^2[U(r)+V(r)\cos^2\theta]+O(a^4),\\
&\! A_{\phi}=\!-\frac{a q_0}{r^2}\sin^2\theta\!+\!O(a^3),\quad A_r=A_\theta=0,\quad\!
q_0=-\frac{R^3}{k},\\
&\! U(r)=-\frac{Ru(R)}{kr^2},  \quad  V(r)=-\frac{3R}{10kr^2}-\frac{q_0}{r^4}-\frac{kq_0^2}{5r^7}.
\end{aligned}
\end{equation}
The metric is expanded consistently through $a^2$ using Eq. \eqref{primaryMetricComponents}.
We evaluate the integrals on a sphere of fixed coordinate radius $r$, with $\xi=\partial_v+\Omega\partial_\phi$ and $\Omega=O(a)$ held fixed in each variation.
Substituting the fields into Eqs. \eqref{fullBoundaryPotential} and \eqref{fullNoetherKernel} reproduces Eq. \eqref{finiteRadiusNoether} and gives
\begin{equation}\label{boundaryCancellation}
\begin{aligned}
\oint\delta\boldsymbol Q_\xi
&=\frac{\delta m}{2}-\Omega\delta(ma)
-\delta\left(\frac{\gamma k^3q_0^3a^2}{6r^8}\right),\\
-\oint\xi\cdot\boldsymbol\Theta
&=\frac{\delta m}{2}
+\delta\left(\frac{\gamma k^3q_0^3a^2}{6r^8}\right).
\end{aligned}
\end{equation}
The omitted terms are $O(a^4)dR+O(a^3)da$.
The $r^{-8}$ contributions cancel between the two integrals, leaving the full surface-charge variation $\delta m-\Omega\delta(ma)$ independent of the extraction radius.
This cancellation holds for arbitrary $u(R)$; the subsequent integrability condition on the normalized horizon charge selects $u(R)$ as described in Sec. \ref{sec:entropy}.

\section{Independent second-order hair variation}\label{app:independenthair}
To test whether the same entropy charge extends to independent hair, write $b=-R^2/k+a^2h_2+O(a^4)$ and treat $(R,a,h_2)$ as independent at fixed $\gamma,k$. 
The geometric horizon conditions still hold through second order. 
The coefficients of $M=M_0+a^2M_2+O(a^4)$ and
$T_H=T_0+a^2T_2+O(a^4)$ become
\begin{equation}
M_2=\frac1{2R}+\gamma kRh_2,\qquad
T_2=-\frac{2+3y}{4\pi R^3}-\frac{3\gamma kh_2}{2\pi R}.
\end{equation}
Using the complete surface charge, with $J=aM_0+O(a^3)$, define $\omega=(dM-\Omega_HdJ)/T_H$. 
Direct differentiation gives
\begin{equation}\label{independentHairCurl}
\partial_{h_2}\omega_R-\partial_R\omega_{h_2}
=\frac{8\pi\gamma kRa^2(1-3y)}{(1+3y)^2}+O(a^4).
\end{equation}
This is nonzero in the domain \eqref{admissibleParameterRegion}.
Thus the same normalized charge cannot define an entropy on the full three-parameter family. 
On the two-parameter subfamily its pullback is integrable for $h_2=1/[k(1-3y)]$, equivalent to Eq. \eqref{exactAmplitudeChoice} at this order. 
A thermodynamic extension with independent hair and an additional work term is a different boundary-variation problem; Eq. \eqref{independentHairCurl} does not exclude it.

%\bibliography{PHreftitle}

\end{document}